\documentclass[reprint,amsmath,amssymb,aip]{revtex4-1}

\usepackage{graphicx}
\usepackage{epstopdf}
\usepackage{amsmath,bm}
\usepackage{threeparttable}
\usepackage{multirow}
\usepackage{booktabs}
\usepackage{color}
\usepackage{hyperref}
\hypersetup{colorlinks=true,linkcolor=blue,filecolor=blue,urlcolor=blue,citecolor=blue,
	}

\begin{document}

\title{Hybrid-Order Topological Phase And Transition in 1$H$ Transition Metal Compounds}%
\author{Ning-Jing Yang}
\affiliation{Fujian Provincial Key Laboratory of Quantum Manipulation and New Energy Materials, College of Physics and Energy, Fujian Normal University, Fuzhou 350117, China}
\affiliation{Fujian Provincial Collaborative Innovation Center for Advanced High-Field Superconducting Materials and Engineering, Fuzhou, 350117, China}

\author{Zhigao Huang}
\affiliation{Fujian Provincial Key Laboratory of Quantum Manipulation and New Energy Materials, College of Physics and Energy, Fujian Normal University, Fuzhou 350117, China}
\affiliation{Fujian Provincial Collaborative Innovation Center for Advanced High-Field Superconducting Materials and Engineering, Fuzhou, 350117, China}

\author{Jian-Min Zhang}
\email[Corresponding author: ]{jmzhang@fjnu.edu.cn}%
\affiliation{Fujian Provincial Key Laboratory of Quantum Manipulation and New Energy Materials, College of Physics and Energy, Fujian Normal University, Fuzhou 350117, China}
\affiliation{Fujian Provincial Collaborative Innovation Center for Advanced High-Field Superconducting Materials and Engineering, Fuzhou, 350117, China}


\begin{abstract}

Inspired by recent experimental observations of hybrid topological states [Hossain, $et \  al.$ Nature, \textbf{628}, 527 (2024)], we predict hybrid-order topological insulators (HyOTIs) in $1H$ transition metal compounds (TMCs), where both second-order and first-order topological states coexist near the Fermi level. Initially, $1H$-TMCs exhibit a second-order topological phase due to the $d$-orbital band gap. Upon coupling of $p$- and $d$- orbitals through the crystal field effect, first-order topological characteristics emerge. This hybrid-order topological phase transition can be tuned via crystal field effects. Combined with first-principles calculations, we illustrate the phase transition with $WTe_2$ and $NbSe_2$. The $WTe_2$ exhibits hybrid-order under ambient conditions, while $NbSe_2$ transitions to hybrid-order under pressure. Additionally, the first-order topological band gap in the HyOTI demonstrates a strong spin Hall effect.
Our findings reveal hybrid-order topological phase in 2D electron materials and underscore spintronic applications.
\end{abstract}

\maketitle

Recently, Shafayat $et \ al.$ discussed the existence of hybrid topological quantum states in solid $\rm As$ in experiment \cite{hossain2024}. By stacking $\rm As$, they observed the coexistence of first-order topological surface and second-order hinge states. 
Hybrid-order topological states generally refer to the coexistence of topological states of different orders within a system.
HyOTIs initially emerged in phononic crystals \cite{zhang2020, PhysRevLett.126.156801}, resulting in dipole and quadrupole moments in different frequency bands. 
In contrast, traditional topological insulators and higher-order topological materials can only host a single type of topological order. For first-order topology, the system will exhibit topological states of dimension d-1, such as surface states in three-dimensional (3D) systems or edge states in two-dimensional (2D) systems \cite{RevModPhys.82.3045,RevModPhys.83.1057,hasan2011three,tokura2019magnetic}. Higher-order topology, on the other hand, manifests topological states of dimension d-2 and higher, such as hinge states and corner states in 3D systems, or corner states in 2D systems \cite{schindler2018higher,xie2021higher,PhysRevLett.127.255501}.
Unlike the bulk-edge correspondence in traditional topological systems \cite{li2023floquet, PhysRevB.108.L121302, PhysRevB.103.195310, PhysRevB.107.085117}, hybrid-order topological materials exhibit distinct correspondences, manifesting as a bulk-surface-edge correspondence in 3D  systems and as a bulk-edge-corner correspondence in 2D systems\cite{ PhysRevLett.126.156801, PhysRevApplied.21.044002}.

However, the bulk-edge-corner correspondence has not yet been explored in 2D electronic materials.
While first-order and higher-order topological materials are quite common, HyOTIs have only been observed in $\rm As$ and $\rm 1T’$-$\rm MoTe_2$ in electronic systems so far \cite{hossain2024, huang2024}. 2D materials offer better tunability and more degrees of freedom compared to 3D materials \cite{zhao2024quantum, tokura2019magnetic, wang2022prediction, cordoba2024topological, wang2021realization, nomoto2023correlation, PhysRevLett.109.266405, levy2020designer}, which suggests that they exhibit richer physical properties than 3D materials \cite{lima2016topologically, PhysRevLett.109.246803, zhang2023topological, liang2024thickness, gao2021layer, bao2022light, lohse2018exploring, strungaru2022ultrafast, fu20232d, PhysRevLett.128.026403}. Additionally, our current understanding of hybrid-order topological phases is limited. Therefore, it is very likely that this type of hybrid-order topological phase is also prevalent in existing 2D materials, possibly even in the most common materials or systems we know \cite{huang2024, PhysRevB.109.115422}. Among the most extensively discussed 2D materials are transition metal compounds (TMCs), which have shown rich physical properties in the fields of superconductivity \cite{cheng2024superconductivity, li2021recent, bawden2016spin, lu2015evidence, kezilebieke2020topological}, valley physics \cite{PhysRevB.104.085149, tyulnev2024valleytronics}, excitonics \cite{sun2022evidence, uddin2022enhanced, zhu2020probing}, spintronics \cite{PhysRevB.99.041101, gorai2016thermoelectricity, pascale2023deeply}, and topology \cite{PhysRevLett.130.116204, PhysRevMaterials.8.044203}. The high-order topological properties of these materials have only been discovered recently \cite{PhysRevLett.130.116204, PhysRevMaterials.8.044203}, highlighting the great physical potential of 2D TMCs.

\begin{figure*}
	\centering
	\includegraphics[width=16cm]{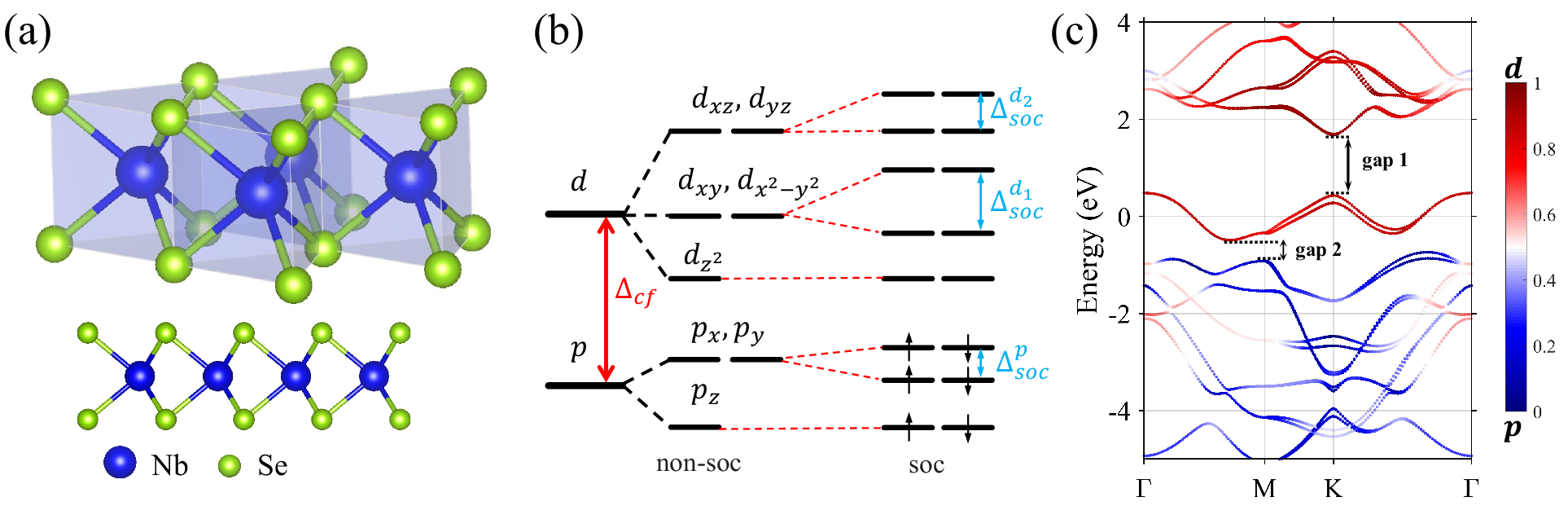}\caption{(a) Crystal structure diagram of $1H$ TMC $\rm NbSe_2$. The blue and green spheres represent Nb and Se atoms, respectively. (b) A schematic diagram of the energy levels occupied by $p$ and $d$ orbitals ($\Delta_{cf}$) in the system. The red arrows represent the crystal field energy level differences between the p- and d-orbitals, while the blue arrows indicate the splitting of other degenerate states caused by SOC. (c) Orbital projected band structure of $\rm NbSe_2$ with considering SOC.}\label{pho:1}
\end{figure*}

In this letter, we identify hidden hybrid-order topological states in $1H$-phase transition metal compounds (TMCs) and elucidate their phase transition process. First, we highlight the second-order topological properties of $1H$-$\rm MX_2$ using the multi-orbital tight-binding (TB) model, which manifest above the Fermi level. Second, a first-order topological phase transition can be induced below the Fermi level by tuning the relative energy levels between the orbitals.
The phase transition involves the conversion between a hybrid-order topological state and a second-order topological state, with contributions from both the $p$ and $d$ orbitals. Finally, we validate the hybrid-order phase transition in $1H$-$\rm MX_2$ under a strain field, using $\rm NbSe_2$ and $\rm WTe_2$ as examples from metals and semiconductors.
The hybrid-order topological states, which switch between first-order and second-order topology near the Fermi level, break the conventional limitations of topological materials. Our results emphasize the importance of hybrid-order topological states in 2D systems and highlight their orbital-dominated nature, providing a pathway for further exploration of hybrid-order quantum materials.

The crystal structure of $1H$-phase TMCs is shown in Fig. \ref{pho:1} (a). In the structural diagram, the blue atoms represent transition metal atoms, while green atoms typically belong to elements from the \uppercase\expandafter{\romannumeral6}A groups. This $1H$ TMC monolayer exhibits the characteristic hexagonal lattice. We take $\rm NbSe_2$ as an example to discuss their common physical properties. The crystal structure of $1H$-$\rm NbSe_2$ belongs to the P-6m2 space group, with its side view depicted in Fig. \ref{pho:1} (a).

For $1H$-phase TMCs, the vicinity of the Fermi level is typically contributed by the $d$ orbitals of the transition metal and the $p$ orbitals of the coordinating elements. Common coordinating elements often come from the oxygen group. The stability of the crystal structure of these TMCs is closely related to the electronic distribution in the $d$ orbitals and the orbital splitting in different crystal fields \cite{kang2021valence, PhysRevB.71.245114, tan2022crucial}. Taking $1H$-$\rm NbSe_2$ as an example. In this compound, the transition metal atom Nb forms bonds with six Se atoms, resulting in a trigonal prismatic crystal structure, as shown in Fig. \ref{pho:1} (a). Under the trigonal prismatic crystal field effect, $d$ orbitals of Nb split into three energy levels: $d_{3z^2-r^2}$, $d_{xy}/d_{x^2-y^2}$, and $d{xz}/d_{yz}$. The $d_{3z^2-r^2}$ orbital occupies a relatively lower energy level, while $d_{xy}/d_{x^2-y^2}$ is slightly higher. The two degenerate $d_{xz}$ and $d_{yz}$ orbitals occupy even higher energy regions. Below these $d$ orbitals, the $p$ orbitals of the Se atoms split into $p_z$ and $p_x/p_y$ energy levels, with $p_z$ being the lowest. The distribution of $p$ and $d$ orbital energy levels in the system is illustrated in Fig. \ref{pho:1} (b). Meanwhile, the spin-orbit coupling of Nb and Se atoms further splits the degenerate energy levels. The energy level splitting of the $d$-orbitals is larger, indicating that the $d$-orbitals play a dominant role in SOC.
Additionally, first-principles calculations of the projected orbital density for $1H$-$\rm NbSe_2$ confirm the localized energy distribution, as shown in Fig. S1. 
The details and discussion of the first-principles calculation parameters can be found in the supplementary materials.

Density functional theory calculations show the orbital projected band structure of $\rm NbSe_2$ in Fig. \ref{pho:1} (c). Since spin-orbit coupling (SOC) is considered, the spin of the energy bands does not exhibit splitting. Unlike $\rm MoS_2$ exhibit metallic behavior due to the $d$ orbital bands crossing the Fermi level. From the projected $p$, $d$ orbital bands, we observe that $d$ orbitals dominate the region above the Fermi level, while below the Fermi level, $p$ orbitals play a significant role. Notably, there is strong coupling between the $p$ and $d$ orbitals, and a clear orbital alternation occurs at the $\Gamma$ point, which has been overlooked in previous studies. This coupling originates from the crystal field effect. The band structure reveals a relatively large gap (gap 1) between the $d$ orbitals, and a smaller gap (gap 2) exists between the $p$ and $d$ orbitals. The crystal field effect associated with the gap 2 between $p$ and $d$ orbitals has been neglected, emphasizing the need to reevaluate the orbital interactions near the Fermi level. We focus on the topological properties of the two global gaps, gap 1 and gap 2, as shown in Fig. \ref{pho:1} (c). TMCs are sensitive to stress fields. In previous studies, the focus was often limited to the physical properties of gap 1, and for simplicity, only the $d$ orbitals were considered, neglecting the role of $p$ orbitals. This led to the oversight of topological properties occurring in the gap 2  interval.

To better analyze the system's topological states and the crystal field effect on hybrid-order topological phase transitions,  we analyze the hidden physical correlations and topological properties using the TB approximation.
We employ a second quantization methods to derive the system's Hamiltonian. This TB model encompasses 5 $d$ orbitals and $2\cdot 3$ $p$ orbitals, where 2 represents the $Se$ atoms in the upper and lower layers, collectively constructing an 11-orbital TB system for $1H$-TMCs. The hoppings in this system are considered only between nearest neighbors. For TMC materials, which often exhibit strong spin-orbit coupling effects, the inclusion of the SOC term in the Hamiltonian is necessary. Therefore, the Hamiltonian for this system can be expressed as:
\begin{equation}
H=\sum_{ i, \alpha} E_{i}^{\alpha}c_{i}^{\dagger}c_{i} + \sum_{ i, j,\alpha,\beta}(t_{i j}^{\alpha,\beta}c_{i\alpha}^{\dagger}c_{j\beta}+h.c.)+\lambda_{s o c}\bm{L}\cdot S ,
\end{equation}
where $c_{i\alpha}^{\dagger}(c_{j\beta})$ represents the electron creation (annihilation) operator for the orbital $\alpha(\beta)$ located at position $i(j)$. $E_i$ represents the onsite energy. $S$ represents the Pauli matrix. $\lambda_{soc}$ denotes the strength of the spin-orbit coupling. Detailed parameters can be found in the supplementary materials.


\begin{figure}
	\centering
	\includegraphics[width=8.5cm]{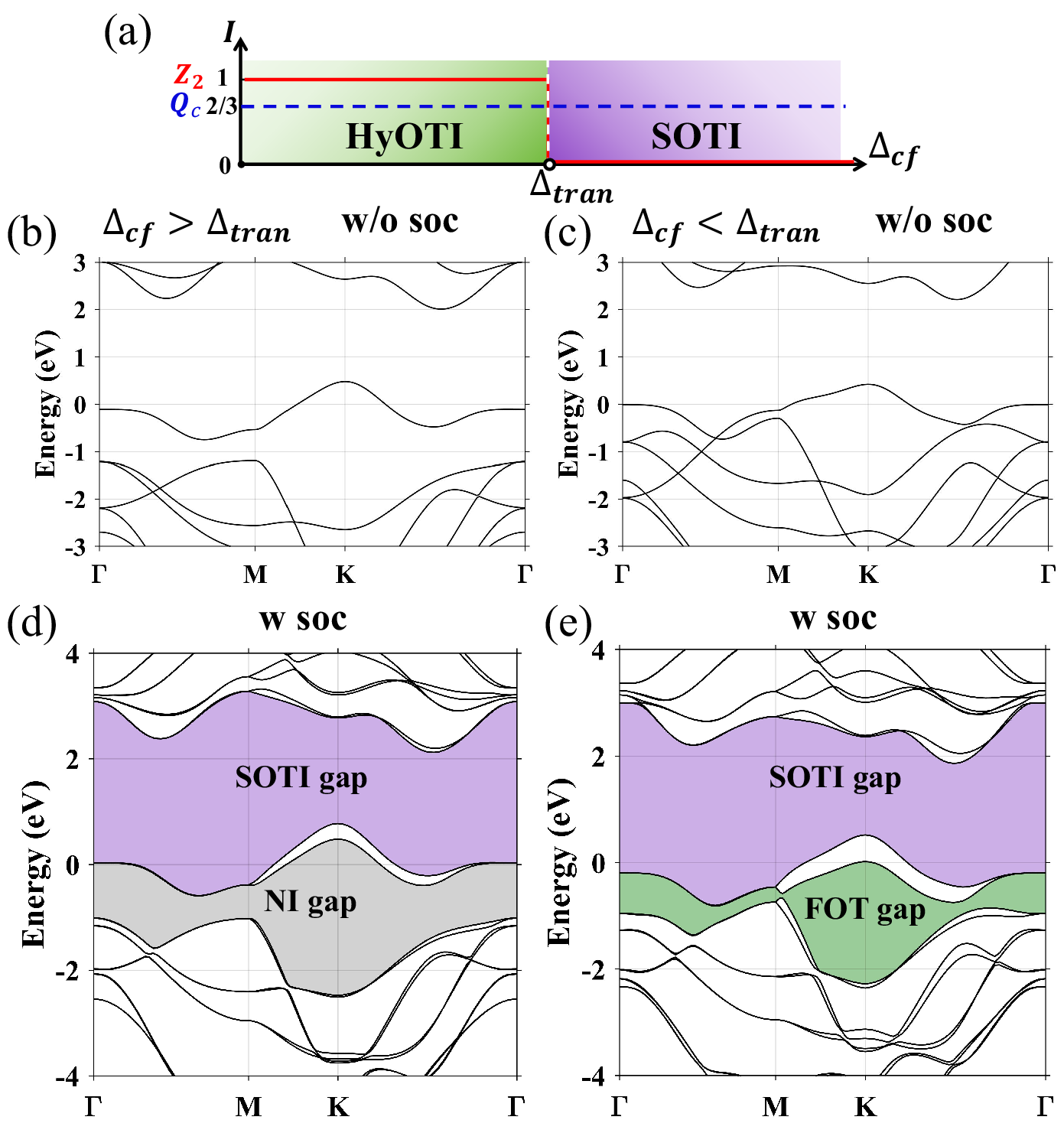}\caption{
			TB model results of the topological phase transition. (a) Diagram of hybrid-Order topological phase transition in $1H$-TMCs. When $\Delta_{cf}$ is larger or smaller than $\Delta_{tran}$, the band structures without SOC correspond to (b, c), respectively. The band results considering SOC are shown in (d, e), highlighting the topological features.}\label{pho:phase}
\end{figure}

To verify the second-order topological insulator (SOTI) of $1H$-TMCs, we first calculate the second order topological indices $Q_c^{(3)}$ of the system, which is protected by the symmetry of the $C_3$ rotation symmetry. For calculating the rotation eigenvalues of all occupied states at the high symmetry points in the Brillouin zone, one can take $[\mathrm{K}_n^{(3)}]=\#\mathrm{K}_n^{(3)}-\#\Gamma_n^{(3)}$, where $\#$ denotes the counting about the symmetry eigenvalues at the points K and $\Gamma$. 
The topological indices \cite{10,11} of the HyOTI are
\begin{equation}
	\chi^{(3)}=([K_1^{(3)}], [K_2^{(3)}]),Q_c^{(3)}=\frac{e}{3}[K_2^{(3)}]\text{mod}\ e,
\end{equation}
where $e$ is the charge of the free electron. 
The upper indicator (3) represents $C_3$ symmetry.  
Then, the topological indicator $\chi^{(3)} = (1, 2) $ and the nonzero corner charge $Q_c^{(3)}=2e/3$ is obtained. 
Thus, we can characterize the second-order topological nature of gap 1 through $Q_c$.
For gap 2 below the Fermi level, the topological property is of the first order. Therefore, for the entire system, we can define a hybrid-order topological index $I=(Q_c, Z_2)$ to characterize the topological property.

\begin{figure}[b]
	\centering
	\includegraphics[width=8.5cm]{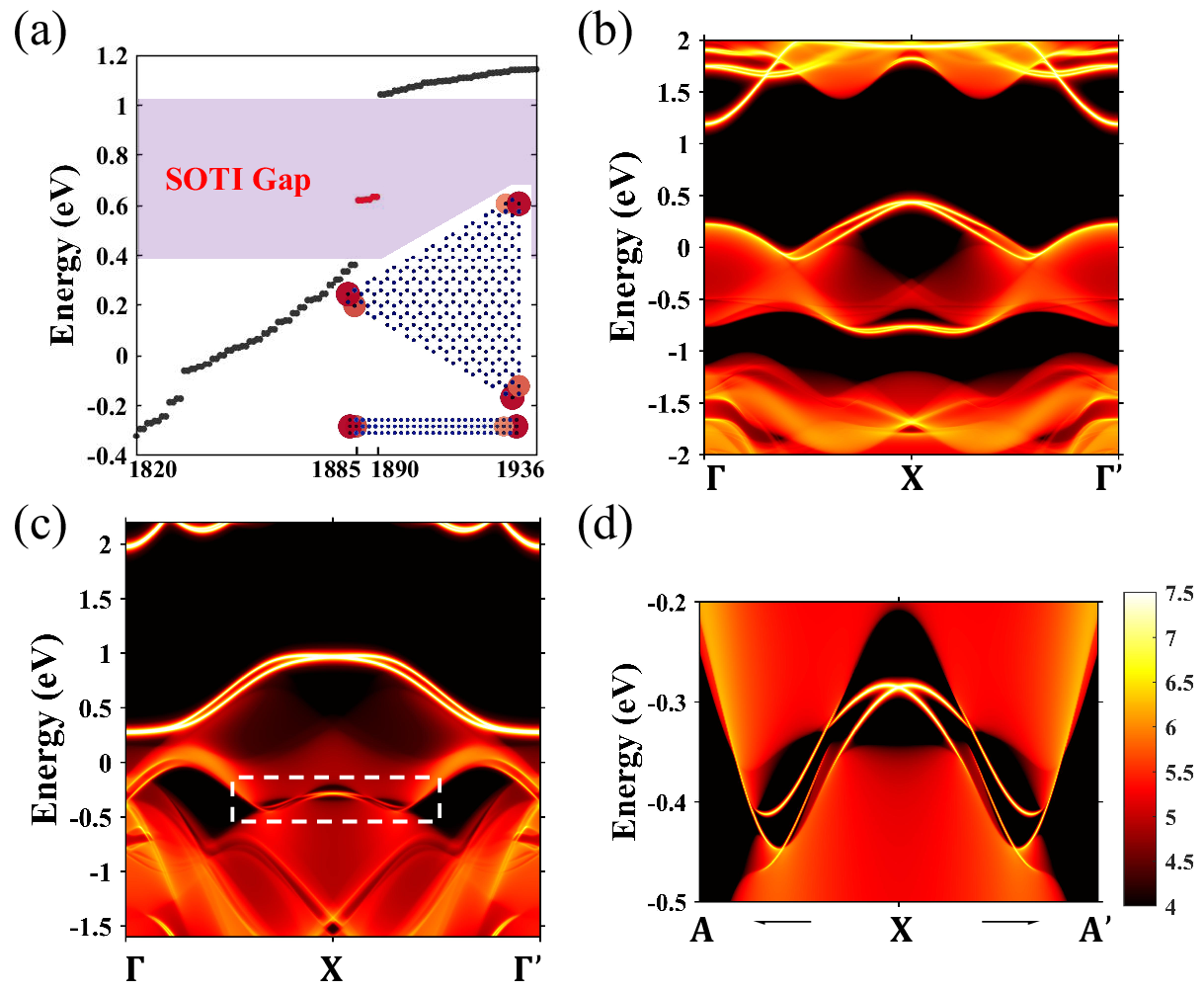}\caption{ The second-order and first-order topological states of $\rm NbSe_2$. (a) Energy spectrum of $1H$-TMCs thin films under SOTI phase. The red dots within the SOTI bandgap represent six corner states. The range of eigenvalue indices for the six corner states is from 1885 to 1890. The localized electronic distribution of angular states occurs at the three corners. (b) Surface band corresponding to the SOTI phase. (c) Surface band structure of the HyOTI phase. The local magnified view within the white dashed box is shown in (d).}\label{pho:surfaceband}
\end{figure}

\begin{figure*}
	\centering
	\includegraphics[width=16cm]{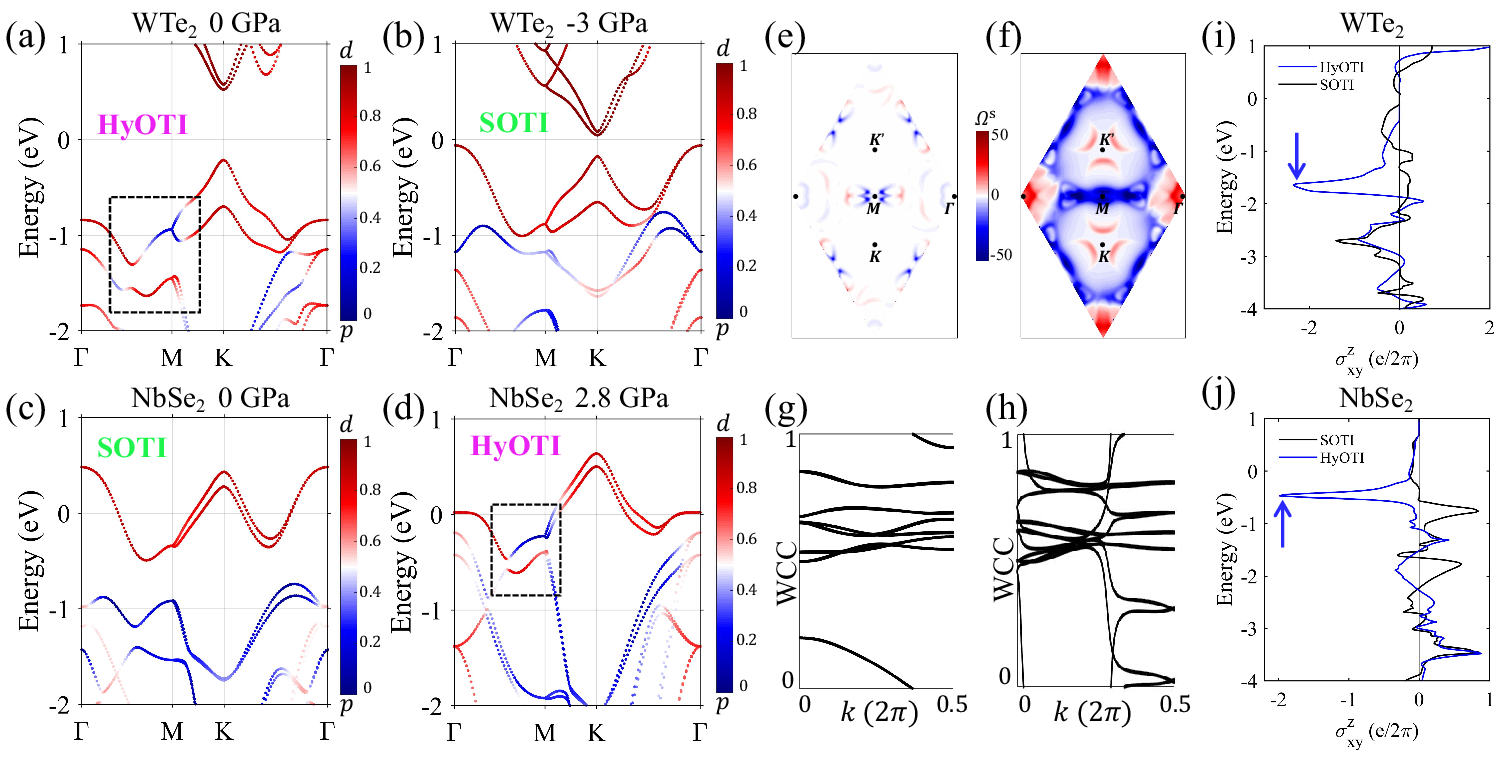}\caption{(a-d) Projected band structures of $WTe_2$ and $NbSe_2$ under different pressures. The spin Berry curvature at the gap 2 energy level for $WTe_2$ is shown in (e) for SOTI and (f) for HyOTI. The WCC characterization of the gap 2 region before and after the topological phase transition is shown in (g) and (h), respectively. The spin Hall conductance (SHC) in SOTI and HyOTI before and after the phase transition are shown in (i) and (j), respectively, as two examples.}\label{pho:examples}
\end{figure*}

Using the multi-orbital TB model, we initially computed the topological properties within the two band gaps. The initial average on-site energy ($E_i$) difference between the $d$ and $p$ orbitals, denoted as $\Delta_{cf}=\overline{E_d}-\overline{E_p}$, is larger than $\Delta_{tran}$, representing the initial state of the system. $\Delta_{tran}$ represents the critical value of the crystal field energy level difference at the phase boundary where the material transitions from HyOTI to SOTI. Through $C_3$ symmetry analysis, the corner charge $Q_c$ is calculated to be $2e/3$, and gap 1 exhibits second-order topological characteristics, marked in purple in Fig. \ref{pho:phase} (d). On the other hand, gap 2 displays trivial insulator properties, indicated by the gray color in Fig. \ref{pho:phase} (d). 
At this time, the system only exhibits second-order topological properties, with $I = (2e/3, 0)$. As $\Delta_{cf}$ decreases, the $p$ orbitals and $d$ orbitals intersect along the high-symmetry line $\Gamma-M$, as shown in Fig \ref{pho:phase} (c). Upon considering SOC, a gap opens at the merged position, and gap 2 exhibits first-order topological (FOT) features with $Z_2$ = 1, as marked in blackish green in Fig. \ref{pho:phase} (e). At this point, the system exhibits a hybrid-order topological state, with $I = (2e/3, 1)$, displaying both second-order and first-order topological characteristics above and below the Fermi level. 
During the $\Delta_e$ modulation process, the critical value for the overlap of the $p$ and $d$ orbitals at the M point is denoted as $\Delta_0$.
At this point, we can obtain two types of topological states near the Fermi level. Due to the constraint of the Fermi level, setting such mixed-order topological states is rare. Currently, they have only been found in TMCs. It can adjust the Fermi level position through gate voltage to enable the switching between different topological orders \cite{lu2015evidence}.

To determine the second-order topological features of $1H$-TMCs, we construct a triangular quantum sheet of $\rm NbSe_2$, as shown in Fig. \ref{pho:surfaceband} (a). The triangular flake, with a side length of 9 unit cells, exhibits six corner states within the SOTI bandgap above the Fermi level. The wave functions of the corner states are localized at the three corner positions, as illustrated in the inset of Fig. \ref{pho:surfaceband} (a). For the intrinsic second-order topological properties of $1H$-TMCs, the surface band structure is shown in Fig. \ref{pho:surfaceband} (b). Two band gaps appear, one above and one below the Fermi level. Notably, the lower band gap is trivial, indicating trivial properties. However, when the system transitions to a HyOTI, the band inversion occurs at the M point, and the interval of the lower band gap exhibits topological properties. The hybrid-order topological surface band structure is depicted in Fig. \ref{pho:surfaceband} (c). By zooming in on the local region in Fig. \ref{pho:surfaceband} (d), it is evident that the topological edge states connect the lower bulk band to the upper bulk band.
Thus, the hybrid-order topological states achieve the bulk-edge-corner correspondence near the Fermi level.


We have not only constructed a multi-orbital TB model for $1H$-TMCs, but also provide specific examples of practical materials. Materials with the potential for hybrid topological order mainly fall into two categories: metallic or semimetallic $1H$-TMCs. For metals, the $d$ orbitals cross the Fermi level, and prominent examples of HyOTIs include ising superconductors $\rm NbSe_2$, $\rm NbTe_2$, $\rm TaSe_2$,  $\rm VTe_2$ and $\rm VSe_2$. As for semimetals, $\rm MX_2 (M = Cr, W, Mo; X = Te, S, Se)$ compounds serve as candidate materials. 
Taking the example of the semimetal $\rm WTe_2$ and the metal $\rm NbSe_2$, we manipulate their phase transitions through tensile and compressive stresses. For $\rm WTe_2$, it exhibits intrinsic hybrid topological properties, as shown in Fig. \ref{pho:examples} (a), where dashed boxes mark the intervals of $p$ and $d$ orbital inversions. When a tensile stress of 3 GPa is applied to $\rm WTe_2$, this $p$, $d$ inversion disappears, as shown in Fig. \ref{pho:examples} (b). Meanwhile, for intrinsic $\rm NbSe_2$, $d$ orbital bands cross the Fermi level, with two band gap spaces above and below the orbital bands, as shown in Fig. \ref{pho:examples} (c). Simultaneously, the calculated corner charge $Q_c = \frac{1}{3}e$ indicates the system’s SOTI characteristics. When a pressure of 2.8 GPa is applied, the $d$ and $p$ orbitals swap positions at the high-symmetry point M, leading to band crossings along the $\Gamma$-M high-symmetry line and the opening of a gap due to SOC. This result perfectly matches the TB model, as depicted in Fig. \ref{pho:examples} (d). 
Due to the first-order topological phase transition to a $Z_2$ topological insulator occurring in the low-energy bandgap, the topological phases before and after the transition can be distinguished by the closely related spin Berry curvature \cite{qe,paoflow}.
Before the $p$, $d$ orbitals inversion, the orbital coupling of gap 2 decreases, resulting in a small spin Berry curvature at that energy level, as shown in Fig. \ref{pho:examples} (e). However, after the $p$, $d$ inversion occurs, there is a significant enhancement in the spin Berry curvature, especially near the $\Gamma$-M high-symmetry line, as depicted in Fig. \ref{pho:examples} (f). This change in spin Berry curvature also reveals the phase transition characteristics of $Z_2$ topological insulators. 
\begin{table}[t]
	\centering
	\caption{Hybrid-order topological materials in $1H$-TMCs}
	\label{tablename}
	
	\vspace{5pt}
	\begin{tabular}{cccc}        
		\hline
		\toprule
		&\multirow{2}{1cm}{Sample} &\multicolumn{2}{c}{HyOTI}  \\
		\cline{3-4}   
		& &Intrinsic & Pressure  \\ 
		\cline{1-4} 
		&Semiconductor &{\begin{tabular}[c]{@{}c@{}}$\rm MoSe_2, MoTe_2$ \\ $\rm WSe_2, WTe_2$ \\ $\rm CrTe_2$ \end{tabular}} 
		&{\begin{tabular}[c]{@{}c@{}}$\rm MoS_2, WS_2$ \\ $\rm CrS_2, CrSe_2$ \end{tabular}} \\		
		\hline
		&Metal &\rm $\rm NbTe_2$ &{\begin{tabular}[c]{@{}c@{}}$\rm NbSe_2$, $\rm TaSe_2$ \\ $\rm VSe_2$, $\rm VTe_2$ \end{tabular}}  \\
		\bottomrule\hline
		
	\end{tabular}\label{sample}
\end{table}

Based on this, by integrating the spin Berry curvature using the spin Hall conductance formula
\begin{equation}
\sigma_{\mathrm{SH}}=\frac{e}{(2\pi)^2}\sum_{n}\int_{\mathrm{BZ}}d^2kf_{n\vec{k}}\Omega_{n,\vec{k}}^{s},
\end{equation}
where $\Omega_{n,\vec{k}}^{s}$ is spin Berry curvature. $f_{n\vec{k}}$ is Fermi distribution function. We can obtain their spin Hall conductance, as shown in Fig. \ref{pho:examples} (i, j). It can be observed that the HyOTI phase exhibits a significantly stronger SHC at the first-order topological bandgap compared to the SOTI phase. 
For gap 2 below the Fermi level, the Wannier charge centers (WCC) under the two different pressures are shown in Fig. \ref{pho:examples} (g) and (h), respectively.
It can be observed that before the $p$ and $d$ orbitals overlap at lower energy levels, gap 2 exhibits a $Z_2$ = 0. After the orbital inversion, gap 2 is characterized by $Z_2$ = 1, indicating topological properties and the system transitions to a hybrid-order topological state.  
Under a unified energy display standard, the differences between SOTI and HyOTI are evident, primarily occurring at the $\Gamma$-M symmetry line and near the $\Gamma$ point.
For the hybrid-order topological candidate materials in the 1$H$-TMCs family, we list several in Table \ref{sample}. Among them, five are intrinsic hybrid-order topological materials. Additionally, six materials exhibit HyOTI topological states only under applied pressure. The DFT calculations for HyOTIs in Table \ref{sample} are presented individually in the supplementary materials.

Although the rule of finding different topological orders in various energy ranges, similar to phononic crystals \cite{zhang2020, PhysRevLett.126.156801}, is relatively straightforward, it is extremely rare to find HyOTIs like those in TMCs, where topological order switching can be achieved at the Fermi level through gate voltage regulation. 
The advantage of gate voltage lies in its non-destructive, reversible, and high-precision control, which does not require altering the material itself. Hybrid-order topology is expected to be realized in multilayer stacked systems through appropriate gate voltage regulation. With different potentials between layers, first-order and second-order topological states can be achieved in the upper and lower layers, respectively.
The topological switchability of HyOTIs breaks the limitations of conventional topological materials.
Moreover, for electronic materials, only topological properties near the Fermi level are meaningful. This is the main reason why we focus our discussion on 1$H$-TMC, as it makes the HyOTI both controllable and significant.
At the same time, although the first-order topology has a non-global band gap, it exhibits a strong spin Hall effect similar to Weyl semimetals \cite{shi2019all}. This characteristic also demonstrates the potential of TMCs in spintronics applications.

In summary, we identify the hybrid-order topological phase and its associated phase transition in $1H$-TMCs from the perspective of multi-orbital coupling. Using a multi-orbital fitted tight-binding model, we highlight the critical role of the $p$-orbital states at low energies in TMCs, which are essential for the hybrid-order topological state. Furthermore, under the influence of the crystal field effect, the hybrid-order topological phase transition in $1H$-TMCs can be readily controlled. We also present the topological phase diagram of $1H$-TMCs as a function of the average energy difference, $\Delta_{cf}$, between the $p$ and $d$ orbitals.
Finally, we confirm the stress-induced hybrid-order topological phase transition in the $1H$ semiconductor $\rm WTe_2$ and the metal $\rm NbSe_2$ through first-principles calculations, highlighting significant changes in spin Hall conductance and the bulk-edge-corner correspondence.
Our results not only broaden the pool of candidate materials for hybrid-order topology but also shed light on the interplay between orbital interactions and the hybrid-order topological state. These findings pave the way for future investigations in hybrid-order topological physics and underscore its promising applications in spintronics.

\vspace{4mm} 
Supplementary Material: ($\mathrm{I}$) Computational details, 
($\mathrm{II}$) Tight-binding model,
($\mathrm{III}$) SHC of Hybrid-Order Topological Phase,
($\mathrm{IV}$) Hybrid-Order Topological Materials in Semiconductors,
($\mathrm{V}$) Hybrid-Order Topological Materials in Metals.

\vspace{4mm} 
We acknowledge useful discussions with Yinhan Zhang. This work is mainly supported by the National Natural Science Foundation of China (No. 11874113) and the Natural Science Foundation of Fujian Province of China (No. 2020J02018).

The authors have no conflicts to disclose.

\textbf{Ning-Jing Yang}: Conceptualization (lead); Data curation (lead); Formal
analysis (lead); Writing – original draft (lead). \textbf{Zhigao Huang}: Funding acquisition
(lead); Project administration (equal); Resources (lead); Supervision
(equal). \textbf{Jian-Min Zhang}: Project administration (lead); Resources
(lead); Supervision (lead); Validation (lead); Visualization (lead);
Writing – review $\&$ editing (lead).

The data that support the findings of this study are available from the corresponding authors upon reasonable request.

\section*{REFERENCES}

\bibliography{References.bib}


\end{document}